\documentclass{aa} 
\usepackage{graphicx}
\usepackage{txfonts}

\begin{document}

\title{{\sc Cinderella}\\Comparison of INDEpendent RELative Least-squares Amplitudes}
\subtitle{Time series data reduction in Fourier Space}

\titlerunning{{\sc Cinderella} -- Time series data reduction in Fourier Space}

\author{P. Reegen\and M. Gruberbauer\and L. Schneider\and W.\,W. Weiss}
\institute{Institut f\"ur Astronomie, Universit\"at Wien, T\"urkenschanzstra\ss e 17, 1180 Vienna, Austria, \email{reegen@astro.univie.ac.at}}

\offprints{P.~Reegen}

\date{Received; accepted}

\abstract
{The identification of increasingly smaller signal from objects observed with a non-perfect instrument in a noisy environment poses a challenge for a statistically clean data analysis.}
{We want to compute the probability of frequencies determined in various data sets to be related or not, which cannot be answered with a simple comparison of amplitudes. Our method provides a statistical estimator for a given signal with different strengths in a set of observations to be of instrumental origin or to be intrinsic.}
{Based on the spectral significance as an unbiased statistical quantity in frequency analysis, Discrete Fourier Transforms (DFTs) of target and background light curves are comparatively examined. The individual False-Alarm Probabilities are used to deduce \em conditional \rm probabilities for a peak in a target spectrum to be real in spite of a corresponding peak in the spectrum of a background or of comparison stars. Alternatively, we can compute joint probabilities of frequencies to occur in the DFT spectra of several data sets simultaneously but with different amplitude, which leads to \em composed \rm spectral significances. These are useful to investigate a star observed in different filters or during several observing runs. The composed spectral significance is a measure for the probability that none of coinciding peaks in the DFT spectra under consideration are due to noise.}
{\sc Cinderella \rm is a mathematical approach to a general statistical problem. Its potential reaches beyond photometry from ground or space: to all cases where a quantitative statistical comparison of periodicities in different data sets is desired. Examples for the composed and the conditional \sc Cinderella \rm mode for different observation setups are presented.}
{}

\maketitle

\section{Introduction}\label{s1} 

The micromag precision, achieved by the MOST\footnote{MOST is a Canadian Space Agency mission, jointly operated by Dynacon Inc., the University of Toronto Institute of Aerospace Studies, the University of British Columbia, and with the assistance of the University of Vienna, Austria.} (Microvariability \& Oscillations of STars) mission (Walker et al.~2003; Matthews 2004), does not only provide exciting new results in asteroseismology, but reveals instrumental problems which challenge our data reduction techniques (see Sect.\,\ref{s1.1}). Cosmic ray impacts on the detector, stray light, positioning errors of the satellite, and thermal stability problems introduce periodic and, in the worst case, pseudo-periodic effects into photometric measurements. All this calls for new techniques in data reduction and analysis (see Sect.\,\ref{s1.2}). 

Space observations in general can provide an unprecedented amount of measurements, requiring an enhanced degree of automatic data analysis without sacrificing accuracy and reliability. In this context, {\sc SigSpec} (Reegen~2007) was developed to combine the Discrete Fourier Transform (DFT) -- a standard method to determine stellar pulsation frequencies -- with a clean statistical quantity: the spectral significance of a peak in an amplitude or power spectrum by comparison to white noise.

The basic idea of {\sc Cinderella} is to use target and comparison data sets simultaneously for a cross-identification of artifacts in the frequency domain. It is the first technique permitting a statistically unbiased and quantitative comparison of different (not necessarily photometric) time series in the {\em frequency} domain. Being applicable to practically all measurements of physical quantities over time, {\sc Cinderella} has the potential to become a valuable tool beyond the scope of micromag space photometry.

\subsection{The MOST mission}\label{s1.1}

The first space telescope designed and built for photometric stellar seismology was EVRIS (Vuillemin et al.~1998), a 10-cm photoelectric telescope aboard the MARS-96 probe, but it unfortunately did not achieve the transfer orbit. An instrument providing photometric information on a large scale useful for asteroseismology was NASA's WIRE satellite, whose primary scientific goal of infrared mapping failed, but a 5-cm star tracker telescope with a CCD detector turned out to permit stellar photometry of remarkable quality (e.\,g., Buzasi et al.~2000). The MOST satellite launched in June, 2003, assumed the role as a precursor to the CNES-led mission COROT (Baglin et al.~2004), which was successfully launched on December 27, 2006, and which is producing extremely useful space photometric data of hitherto unprecedented accuracy and volume.

MOST, WIRE and COROT are low-Earth-orbit (LEO) missions with comparable environmental effects (e.g., cosmic radiation, stray light scattered from the Earth's surface). A further commonality of all three missions is the requirement to extract asteroseismic information from a series of up to hundreds of thousands of CCD frames (or sub-rasters, respectively), each of which may consist of a few hundred to several million pixels. Hence, the present work may apply to other LEO space photometry missions and to ground-based multi-object photometry.

The MOST telescope is a 15-cm Maksutov optical telescope, supplied with a single broadband filter and initially with two identical CCD detectors: one used for science data acquisition, the other for the {\em Attitude Control System} (ACS). Thanks to the low mass of $54$\,kg and the ACS developed by Dynacon, Inc. (Groccott, Zee \& Matthews~2003; Carroll, Rucinski \& Zee~2004), a pointing stability to approximately $\pm 1\arcsec$ rms is achieved.

In {\em Fabry Imaging} mode the telescope entrance pupil is imaged onto the CCD via a Fabry microlens as is shown by Figs.\,7 and 8 of Walker et al.~(2003). Each Fabry Image is an annulus with an outer diameter of 44 pixels. The pixels in a square subraster outside the annulus are used to estimate the background. MOST also obtains {\em Direct Imaging} photometry of typically $1-6$ stars, based on defocussed images (FWHM $\sim$ 2.2 pixels; Rowe et al.~2006; Huber \& Reegen~2008), and {\em Guide Star} photometry of about $20-30$ stars (Aerts et al.~2006; Saio et al.~2006).

\subsection{Data reduction}\label{s1.2}

The data reduction described by Reegen et al.~(2006) applies linear correlations between pairs of target and background pixels for stray light correction. This so-called {\em decorrelation technique} is also applicable to simultaneous photometry of several stars, in this case correlating variable vs.~constant stars.

\begin{figure}\includegraphics[width=256pt]{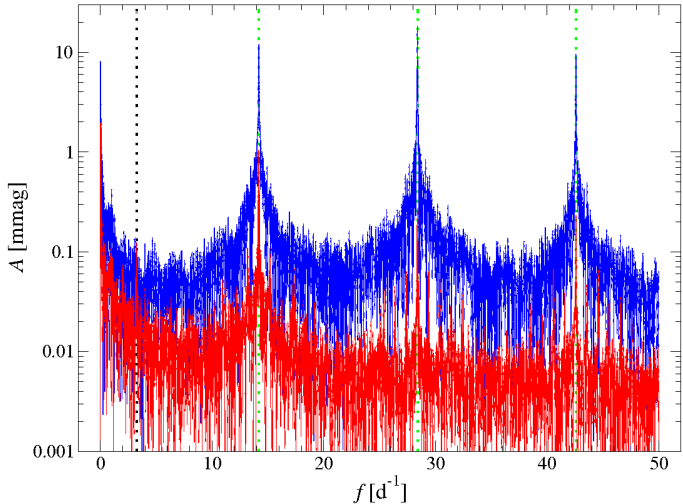}
\caption{The raw light curve ({\em blue}) of the MOST Fabry target $\beta$\,CMi  and after data reduction ({\em red}).  Harmonics of the satellite's orbital frequency ($\approx 14.2$\,d$^{-1}$; {\em dotted green}), the detected stellar signal ($3.257$\,d$^{-1}$ \& $3.282$\,d$^{-1}$; {\em dotted black}) are indicated.}\label{fig1}
\end{figure}

Fig.\,\ref{fig1} illustrates the performance of the Fabry imaging photometry with MOST data of $\beta$ CMi (Saio et al.~2007). The blue graph refers to the raw data and the red graph to the reduced light curve. The overall noise level decreased by an order of 10, and so did the harmonics of the orbital frequency of the spacecraft, ($\approx 14.2$\,d$^{-1}$ for $101.4$\,min; Walker et al.~2003). However, instrumental peaks (dotted green lines) persisted on a lower level and their amplitudes still exceeded the stellar signal (main frequencies: $3.257$\,d$^{-1}$ \& $3.282$\,d$^{-1}$; dotted black line).

\subsection{\sc SigSpec}\label{s1.3}

{\sc SigSpec} (Reegen~2007), is based on DFT amplitude spectra and consecutive prewhitening of dominant peaks. But instead of considering the peak with the highest amplitude to be significant and estimating the reliability roughly in terms of signal-to-noise ratio, the {\em Probability Density Function} (PDF) is employed. The PDF depends on the frequency and phase of the examined peak using white noise as a reference. The mean photometric magnitude in a time series is usually reduced to zero before evaluating the DFT. {\sc SigSpec} the resulting statistical consequences into account, and is furthermore not restricted to Gaussian distributed residuals.

The {\em False-Alarm Probability} is a frequently used statistical quantity in time series analysis. It is the probability of a peak at a given amplitude level to be generated by noise. Formally it is obtained through integration of the PDF. To avoid problems in computing extremely low numerical values, {\sc SigSpec} returns a quantity called {\em spectral significance} (hereafter abbreviated by ``sig''), which is the negative logarithm of the False-Alarm Probability. It gives  the number of uncorrelated data sets needed, containing pure noise, so that a peak in the Fourier domain appears which is comparable in amplitude and phase to the peak under consideration in the observed data.

Although {\sc SigSpec} prevailed as a powerful tool for analyzing MOST photometry, it occasionally suffered from the weakness of having to refer to uncorrelated (i.e. white) noise. 

\subsection{The virtue of {\sc Cinderella}}\label{s1.4}

Frequencies with individual amplitudes and phases (``peaks'') in the DFT spectra of a target and comparison data sets are examined by {\sc Cinderella} for compatibility. In other words, {\sc Cinderella} allows us to investigate whether these data sets are related by any physical (deterministic) process. The procedure is the same if the comparison data represent sky background or a star with a different frequency spectrum as the target star, which -- in the best case -- is a constant star. Subsequently, the terms ``target star'' and ``comparison star'' will be used, keeping in mind that everything discussed here readily applies to sky readings instead of comparison stars as well. Obviously, all compared data sets have to be observed under similar circumstances. An extension of the method to handle more than one comparison data set is useful for multi-object environments, such as  photometry in a field.

In conditional mode, {\sc Cinderella} establishes a quantitative comparison of significant frequencies occurring at the same time in at least two different data sets. It returns a statistically robust value, called {\em conditional sig}, for the probability that a peak in the spectrum of one data set is not (deterministically) related to a peak in the other data set(s) within a given frequency resolution.

The alternative composed mode is dedicated to testing whether peaks in different DFT spectra with similar frequencies are ``real'', in the sense of not due to noise. The corresponding quantity, the {\em composed sig} is a measure for the probability that none of the examined peaks is due to noise.

\subsection{Frequency resolution}\label{s1.6}

The question how to set the frequency difference acceptable for the consideration of peaks as coincidental is crucial to the examination of corresponding peaks in different DFT spectra. In this context, an alternative definition to the Rayleigh resolution,
\begin{equation}\label{eqRayleigh}
\delta f_R := \frac{1}{\Delta t}\: ,
\end{equation}
with $\Delta t$ denoting the total time interval width of the time series is introduced by Kallinger, Reegen \& Weiss~(2007). They suggest to additionally employ the sig for a peak amplitude according to
\begin{equation}\label{eqKallinger}
\delta f_K := \frac{1}{\Delta t\sqrt{\mathrm{sig}\left( A\right)}}\: ,
\end{equation}
for obtaining a more realistic criterion for matching peaks ({\em frequency resolution}) than provided by Eq.\,(\ref{eqRayleigh}). Their numerical simulations show an excellent compatibility of this quantity, subsequently termed {\em Kallinger resolution}, to the {\em frequency error} derived by Montgomery \& O'Donoghue~(1999).

For practical applications, it is useful to enhance the flexibility of {\sc Cinderella} by introducing an exponent $z$ and to re-define the frequency resolution according to
\begin{equation}\label{eqFRes}
\delta f := \frac{1}{\Delta t\left[\sqrt{\mathrm{sig}\left( A\right)}\right] ^z}\: ,
\end{equation}
where $z$ usually attains values in the range $\left[ 0,1\right]$. The Rayleigh resolution is obtained for $z=0$, whereas $z=1$ yields the Kallinger resolution.

\section{Theory}\label{s2}

The theoretical framework of {\sc Cinderella} presented here contains a conversion that makes amplitudes in the DFT spectra of different datasets comparable, introduces conditional and composed sig, discusses how to handle peaks in a target dataset without a corresponding counterpart in the comparison dataset, and generalizes the method to multiple comparison datasets.

\subsection{Amplitude transformation between different mean magnitudes}\label{s2.1}

Assuming that stray light artifacts are additive in terms of intensity, a signal amplitude detected in a comparison data set may readily be inherited for a comparison with the target amplitude, if intensities were employed for the frequency analysis. The corresponding magnitude variations appear on a scaling that depends on the average magnitude. This is reasonable for instrumental effects as well. Let us further assume that mean intensities $\left< I\right>$ are converted into mean magnitudes $\left< m\right>$ according to
\begin{equation}\label{eq1}
\left< m\right> = -2.5\log\left< I\right>\: ,
\end{equation}
which holds to a sufficient approximation if the variations are small compared to the mean intensity. In strict terms, a geometrical mean intensity transforms in to an arithmetical mean magnitude.

Given a mean magnitude $\left< m_C\right>$ and a stray-light induced sinusoidal variation with amplitude $A_C$ (in magnitudes), the maximum intensity in the comparison light curve will be
\begin{equation}\label{eq2}
\left< I_C\right> + \Delta I = 10^{-0.4\left(\left< m_C\right> - A_C\right)}\: ,
\end{equation}
where $\left< I_C\right>$ denotes the mean intensity of the comparison data and $\Delta I$ is the intenstity amplitude corresponding to $A_C$. Thus an estimate of the intensity amplitude is obtained by
\begin{equation}\label{eq3}
\Delta I = 10^{-0.4\left(\left< m_C\right> - A_C\right)} - 10^{-0.4\left< m_C\right>}\: .
\end{equation}

This magnitude-intensity transformation of amplitudes uses the maximum and mean intensities only. The reason is that variations are distorted by the logarithmic scaling, and this distortion is stronger towards low intensities. Hence the Gaussian error propagation (producing symmetric errors only) is not appropriate, nor is it advisable to encounter the minimum intensity as an estimator. Both statements were confirmed by numerical simulations.

Since the stray-light induced variation is assumed additive in terms of intensity, the maximum target intensity will be
\begin{equation}\label{eq4}
\left< I_T\right> + \Delta I = 10^{-0.4\left< m_T\right>} + 10^{-0.4\left(\left< m_C\right> - A_C\right)} - 10^{-0.4\left< m_C\right>}\: ,
\end{equation}
substituting for $\Delta I$ according to Eq.\,(\ref{eq3}). The approximation
\begin{equation}\label{eq5}
A_T \approx 2.5\log\left( 1 + \frac{\Delta I}{\left< I_T\right>}\right)
\end{equation}
for the target amplitude corresponding to a comparison amplitude $A_C$ leads to
\begin{equation}\label{eq6}
A_T \approx 2.5\log\left[ 1 + \frac{10^{-0.4\left(\left< m_C\right> - A_C\right)} - 10^{-0.4\left< m_C\right>}}{10^{-0.4\left< m_T\right>}}\right]\: .
\end{equation}

This is an estimator of the amplitude in a target star corresponding to artificial intensity variations of amplitude $A_C$ in a comparison star.

At this point it has to be emphasized that this is a theoretically consistent transformation that will yield a reasonable estimate in many practical applications. However, the detailed study of contaminated measurements may occasionally demand special approaches to the calibration of magnitudes. Such an example is presented in Sect.\,\ref{s3} and discussed in detail therein.

\subsection{Frequency and phase differences}\label{s2.2}

If a peak in the DFT amplitude spectrum of a comparison dataset is found within the Rayleigh or Kallinger frequency resolution, respectively, about a target peak, the two considered frequencies and phases generally do not match perfectly. We know that DFT peak amplitudes show systematic deviations for different frequencies and phases (e.\,g. Kovacs 1980), whence a transformed amplitude $A_T$ at a frequency $\omega ^\prime$ and a phase angle $\theta ^\prime$ in Fourier Space need not refer to the same amplitude at the frequency $\omega$ and the phase angle $\theta$ of the corresponding target peak. However, since all calculations were performed using {\sc SigSpec} (Reegen~2007) and since the amplitudes are optimized by least-squares fits, they may be considered free of such effects to a satisfactory extent.

At the present status of our investigations, we omit possible effects of frequency and phase lag. Under the condition of the same instrumental or environmental process to be responsible for both target and comparison signal, the frequencies are expected to be equal. In addition, frequency deviations are already taken into account for candidate selection. This is why the frequencies in the target and comparison data are considered equal at this stage of calculation. On the other hand, it was pointed out by Reegen et al.~(2006) that stray light moving over a detector produces phase differences in the stray light signal measured at different positions on the CCD. These phase lags are the main constraint to the quality of the data reduction procedure described there. Hence it definitely makes sense to omit the phase information in the technique introduced here and consider all signal phases consistently aligned to the phase in the target dataset.

\subsection{Conditional spectral significance}\label{s2.3}

The interesting question is now, {\em ``What is the probability that a given target peak with an amplitude $A\left(\omega ,\theta\right)$ is generated by the same process as a transformed comparison peak with an amplitude $A_T\left(\omega ^\prime ,\theta ^\prime\right)$?''} The answer may be given in terms of sig.

According to Sect.\,\ref{s2.1}, we may use $A_T\left(\omega ,\theta\right) \approx A_T\left(\omega ^\prime ,\theta ^\prime\right)$. If a comparison of sigs is desired for constant time-domain sampling, frequency and phase, then the calculations simplify to a comparison of signal-to-noise ratios,
\begin{equation}\label{eq7}
\frac{\mathrm{sig}\left(A,\omega, \theta\right)}{\mathrm{sig}\left( A_T,\omega ^\prime ,\theta ^\prime\right)} = \left(\frac{A}{A_T}\right) ^2\frac{\left< x_T^2\right>}{\left< x^2\right>}\: ,
\end{equation}
where $\left< x^2\right>$ denotes the variance of the target dataset including the signal itself, and $\left< x_T^2\right>$ is the variance the target dataset would have if the amplitude were $A_T$ instead of $A$. Annotating the variance of the target light curve after prewhitening $\left< x_P^2\right>$, the scaling from $A$ onto $A_T$ is obtained via the difference of variances $\left< x^2\right> - \left< x_P^2\right>$, which is a measure for the amount of signal prewhitened for an amplitude $A$. If an amplitude $A_T$ is used instead, the corresponding amount will transform into $\left(\frac{A_T}{A}\right) ^2\left(\left< x^2\right> - \left< x_P^2\right>\right)$. Then the variance $\left< x_T^2\right>$ immediately evaluates to
\begin{equation}
\left< x_T^2\right> = \left< x_P^2\right> + \left(\frac{A_T}{A}\right) ^2\left(\left< x^2\right> - \left< x_P^2\right>\right)\: .
\end{equation}
This expression transforms Eq.\,(\ref{eq7}) into
\begin{equation}\label{eq8}
\frac{\mathrm{sig}\left(A,\omega, \theta\right)}{\mathrm{sig}\left( A_T,\omega ^\prime ,\theta ^\prime\right)} = 1 + \left[\left(\frac{A}{A_T}\right) ^2 - 1\right]\frac{\left< x_P^2\right>}{\left< x^2\right>}\: .
\end{equation}

The conditional False-Alarm Probability of producing at least an amplitude $A$, if an amplitude $A_T$ is presumed, is a fraction of the corresponding individual False-Alarm Probabilities, if the corresponding processes are independent. The sig is defined as the (negative) logarithm of False-Alarm Probability, whence a ratio of False-Alarm Probabilities corresponds to a difference of sigs, i.\,e., we obtain
\begin{equation}\label{eq10}
\mathrm{sig}\left( A\left|\right. A_T,\omega ,\theta\right) = \mathrm{sig}\left( A,\omega ,\theta\right)\left\lbrace 1\! - \frac{1}{1\! + \left[\left(\frac{A}{A_T}\right) ^2\! - 1\right]\frac{\left< x_P^2\right>}{\left< x^2\right>}}\right\rbrace\: .
\end{equation}
This is the {\em conditional sig} of a target peak with an amplitude $A$ under consideration of a comparison peak with a transformed amplitude $A_T$, where the transformation of the comparison amplitude may be performed according to Eq.\,(\ref{eq6}). E.\,g., a peak with a conditional sig of 2 is true despite the given comparison peak in 99 out of 100 cases.

The computation of conditional sigs for multiple comparison datasets contains the {\sc Cinderella} analysis of the target dataset under consideration vs.~each individual comparison dataset. Then the individual conditional sigs may be averaged over all comparison datasets. The resulting mean conditional sig and the corresponding rms error are reasonable estimators for the overall reliability of a target peak. In practical applications, one will trust in a target peak if the mean conditional sig is high, both in absolute numbers and in units of rms error.

\subsection{Joint distributions}\label{s2.4}

An alternative question, relevant in some cases, is, {\em ``Given two independently measured datasets, what is the (joint) probability of a coincident peak not to be due to noise in both datasets?''} The difference with respect to Sect.\,\ref{s2.3} is that here none of the two time series is treated as a mere comparison dataset. This question may, e.\,g., apply to differential photometry of the same target with respect to different comparison stars, or to measurements of the same target in different years. The considered case refers to a logical `and'.

Given two statistically independent time series with two coincident peaks at sigs $\mathrm{sig}\left( A_1\right)$, $\mathrm{sig}\left( A_2\right)$, the False-Alarm Probability, $\Phi _{\mathrm{FA}\,1,2} = 10^{-\mathrm{sig}\,\left( A_{1,2}\right)}$ of an individual peak is the probability that it is generated by noise. The complementary probability that the considered peak is true is $1 - 10^{-\mathrm{sig}\,\left( A_{1,2}\right)}$. If the individual components are statistically independent, the (joint) probability of all components to be real is the product of the individual probabilities,
\begin{equation}\label{eq11}
1 - \Phi _{\mathrm{FA}} = \left(1 - \Phi _{\mathrm{FA}\,1}\right)\left(1 - \Phi _{\mathrm{FA}\,2}\right)\: .
\end{equation}
Consistently, a ``joint sig'' is introduced as the negative logarithm of the total False-Alarm Probability, $\Phi _{\mathrm{FA}}$, and in terms of individual sigs, one obtains
\begin{equation}\label{eq12}
\mathrm{sig}\left( A_1\wedge A_2\right) := - \log\left\lbrace 1 - \left[ 1 - 10^{-\mathrm{sig}\,\left( A_1\right)}\right]\left[ 1 - 10^{-\mathrm{sig}\,\left( A_2\right)}\right]\right\rbrace\, .
\end{equation}

In computational applications, numerical problems may come along with a straight-forward implementation of this relation, namely if $10^{-\mathrm{sig}\,\left( A_i\right)}$ produces an overflow. If $\mathrm{sig}\left( A_2\right)$ is high and $\mathrm{sig}\left( A_1\right) > \mathrm{sig}\left( A_2\right)$, then the resulting joint sig will be $\mathrm{sig}\left( A_1\wedge A_2\right) \approx \mathrm{sig}\left( A_2\right)$, and the amount of change in $\mathrm{sig}\left( A_2\right)$ by the composition with $\mathrm{sig}\left( A_1\right)$ may be calculated by a linear estimate according to
\begin{equation}\label{eq13}
\mathrm{sig}\left( A_1\wedge A_2\right)\approx\mathrm{sig}\left( A_2\right) +\left.\frac{d\,\mathrm{sig}\left( A_1\wedge A_2\right)}{d\,\Phi _{\mathrm{FA}\,1}}\right| _{\Phi _{\mathrm{FA}\,1} = 0}\Phi _{\mathrm{FA}\,1}\: ,
\end{equation}
which evaluates to
\begin{equation}\label{eq14}
\mathrm{sig}\left( A_1\wedge A_2\right)\approx\mathrm{sig}\left( A_2\right) -\left(\frac{1}{\Phi _{\mathrm{FA}\,2}}-1\right)\Phi _{\mathrm{FA}\,1}\log e\: .
\end{equation}
For $\Phi _{\mathrm{FA}\,2}\ll 1$, we may set $\frac{1}{\Phi _{\mathrm{FA}\,2}}-1\approx\frac{1}{\Phi _{\mathrm{FA}\,2}}$, which yields
\begin{equation}\label{eq15}
\mathrm{sig}\left( A_1\wedge A_2\right)\approx\mathrm{sig}\left( A_2\right) -10^{\mathrm{sig}\,\left( A_2\right) - \mathrm{sig}\,\left( A_1\right)}\log e\: .
\end{equation}
If $\mathrm{sig}\left( A_1\right)$, $\mathrm{sig}\left( A_2\right)$ differ by e.\,g.~5, the joint sig will deviate from $\min\left[\mathrm{sig}\left( A_1\right),\mathrm{sig}\left( A_2\right)\right]$ in the 5th digit.

If more than two, say $N$, time series are examined, Eq.\,(\ref{eq12}) may be generalized to
\begin{equation}\label{eq16}
\mathrm{sig}\left(\bigwedge A_n\right) := - \log\left\lbrace 1 - \prod_{n=1}^{N}\left[ 1 - 10^{-\mathrm{sig}\,\left( A_n\right)}\right]\right\rbrace\: .
\end{equation}

In practical applications, the employment of the joint sig as an estimator for the reliability of a peak in several different DFT spectra simultaneously may lead to very low absolute sig values. This becomes evident, if we consider $N$ corresponding peaks at the same sig level $\mathrm{csig}$. Then the composed sig evaluates to
\begin{equation}\label{eq17}
\mathrm{sig}\left(\bigwedge A_n\right) = - \log\left[ 1 - \left( 1 - 10^{-\mathrm{csig}}\right) ^N\right]\: ,
\end{equation}
which consistently decreases with increasing number of datasets $N$.

Setting $\mathrm{csig} =: \frac{\pi}{4}\log e$, which is the expected sig for white noise (Reegen~2007), Eq.\,(\ref{eq16}) evaluates to
\begin{equation}\label{eq18}
\mathrm{sig}\left(\bigwedge A_n\right) = - \log\left\lbrace 1 - \left[ 1 - \exp\left( -\frac{\pi}{4}\right)\right] ^{\,N}\right\rbrace\: .
\end{equation}
This makes clear that both the sigs of given peaks as well as the ``noise'' in the significance spectrum will consistently decrease with the number of employed time series.

\subsection{Composed spectral significance}\label{s2.5}

The dependence of the statistical properties of the joint sig on the number of datasets is potentially irritating, since it does not provide numerical values that can be interpreted at first glance. Thus it is convenient to introduce a more intuitive scaling.

Eq.\,(\ref{eq17}) may be re-written as
\begin{equation}\label{eq19}
\mathrm{csig}\left( A_n\right) = - \log\left[ 1 - \sqrt[N]{1 - 10^{-\mathrm{sig}\,\left(\bigwedge A_n\right)}}\,\right]\: ,
\end{equation}
where $\mathrm{csig}$ -- the {\em composed sig} -- is now considered as a function of $A_n$. The meaning becomes transparent substituting Eq.\,(\ref{eq16}) for $\mathrm{sig}\left(\bigwedge A_n\right)$, which yields
\begin{equation}\label{eq20}
\mathrm{csig}\left( A_n\right) = - \log\left\lbrace 1 - \sqrt[N\,]{\prod_{n=1}^{N}\left[ 1 - 10^{-\mathrm{sig}\,\left( A_n\right)}\right]}\,\right\rbrace\: .
\end{equation}
The composed sig of a sample of corresponding peaks is the unique sig level for the individual peaks that would reproduce the given joint probability. The advantage of this quantity is that it is essentially independent of the number of datasets under consideration.

\subsection{Trust coefficient}\label{s2.6}

A related question is, {\em ``Given $N$ datasets and an associated composed sig for a set of corresponding peaks therein, what is the fraction of datasets in which the considered peak is significant?''} Since sig is a floating-point number rather than a binary output in the sense of, ``This peak is true/false'', it does not provide a unique basis for the decision whether to consider a given peak due to noise. But if we {\em assign} two constant sig levels $a$, $r$ to acceptance and rejection of a peak, respectively, Eq.\,(\ref{eq16}) may be written as
\begin{equation}\label{eq21}
\mathrm{sig}\left(\bigwedge A_n\right) = - \log\left[ 1 - \left( 1 - 10^{-a}\right) ^M\left( 1 - 10^{-r}\right) ^{N-M}\right]\: ,
\end{equation}
if $M$ out of the $N$ peaks are accepted. Expressing this relation in terms of $\tau _r^a := \frac{M}{N}$, we obtain
\begin{equation}\label{eq22}
\tau _r^a = \frac{\frac{1}{N}\log\left[ 1-10^{-\mathrm{sig}\left(\bigwedge A_n\right)}\right] - \log\left( 1 - 10^{-r}\right)}{\log\left( 1 - 10^{-a}\right) - \log\left( 1 - 10^{-r}\right)}\: ,
\end{equation}
for the fraction of accepted peaks in the examined sample. The function $\tau$ is called the {\em trust coefficient}. It is the fraction of reliable peaks in a sample of $N$ datasets, based on the assumed sig levels $a$ for an accepted peak (not due to noise) and $r$ for a rejected peak (due to noise).

Substituting $\mathrm{sig}\left(\bigwedge A_n\right)$ by the right-hand expression in Eq.\,(\ref{eq16}) transforms Eq.\,(\ref{eq22}) into
\begin{equation}\label{eq23}
\tau _r^a = \frac{\frac{1}{N}\left\lbrace\sum _{n=1}^N \log\left[ 1-10^{-\mathrm{sig}\left( A_n\right)}\right]\right\rbrace - \log\left( 1 - 10^{-r}\right)}{\log\left( 1 - 10^{-a}\right) - \log\left( 1 - 10^{-r}\right)}\: .
\end{equation}
On the other hand, the trust coefficient is related to the composed sig via
\begin{equation}\label{eq24}
\tau _r^a = \frac{\log\left[ 1-10^{-\mathrm{csig}\,\left( A_n\right)}\right] - \log\left( 1 - 10^{-r}\right)}{\log\left( 1 - 10^{-a}\right) - \log\left( 1 - 10^{-r}\right)}\: ,
\end{equation}
which follows from Eqs.\,(\ref{eq20}) and (\ref{eq23}). Since the composed sig is independent of the number of examined spectra, the trust coefficient is independent as well.

\begin{figure}\includegraphics[width=256pt]{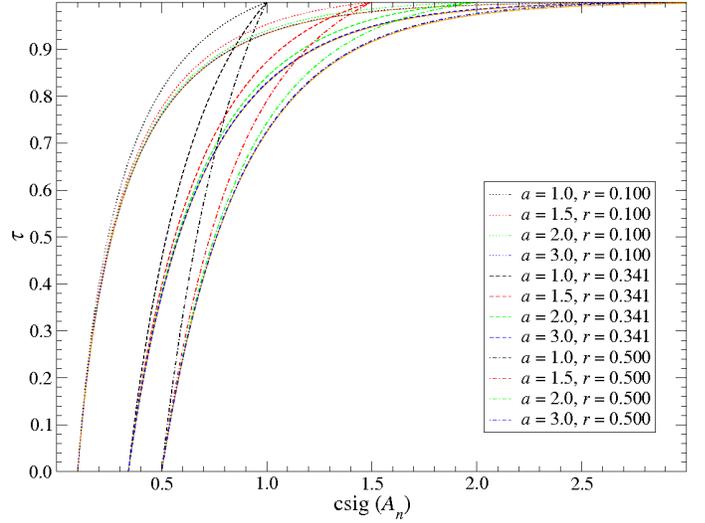}
\caption{Relation between the trust coefficient and the composed sig; $12$ different constellations of the constant sig levels assigned for acceptance ($a$) and rejection ($r$) are displayed. The {\em orange} lines represent the solutions for $a\to\infty$.}
\label{fig:trust}
\end{figure}

Fig.\,\ref{fig:trust} displays the relation between the trust coefficient and the composed sig for altogether $12$ parameter combinations where $a\in\left\lbrace 1,1.5,2,3\right\rbrace$ and $r\in\left\lbrace 0.1,\frac{\pi}{4}\log e,0.5\right\rbrace$. For $\mathrm{csig}\left( A_n\right) < r$, the trust coefficient is $0$, for $\mathrm{csig}\left( A_n\right) > a$ it is $1$. Furthermore, for $a\to\infty$, Eq.\,(\ref{eq24}) yields
\begin{equation}\label{eq25}
\tau _r^\infty = 1 - \frac{\log\left[ 1-10^{-\mathrm{csig}\,\left( A_n\right)}\right]}{\log\left( 1 - 10^{-r}\right)}\: ,
\end{equation}
which is indicated by the {\em orange} lines in the figure. For all three values of $r$, the graphs for $\tau _r^3$ and $\tau _r^\infty$ are practically identical. Thus, $\tau _{\frac{\pi}{4}\log e}^\infty$ will provide a reasonable estimator for the percentage of significant peaks in a sample in practical applications.

\subsection{Peaks without coincidences}\label{s2.7}

The search and comparison of coincident peaks raises the question how to treat signal components that have no counterpart in the comparison spectra. According to our present practical experience, it is in such cases reasonable to assign a constant sig level of $\frac{\pi}{4}\log e \approx 0.341$ (the expected sig for white noise) to the comparison data. Then a target peak, for which no significant coincidence is detected, can be compared to the expected value for pure noise by default.

\section{Conditional spectral significance applied to MOST photometry}\label{s3} 

In Sect.\,\ref{s2.3}, the conditional sig was introduced as a measure of the probability that a specific peak in a DFT spectrum (characterized by frequency, amplitude and phase) is deterministically linked to a peak in another dataset within the frequency resolution (Eq.\,(\ref{eqFRes})). Considering one of the two datasets to represent the sky background or a constant comparison star, this concept can be used to isolate intrinsic frequencies from instrumental or environmental periodicities. If a peak in the target data has a significant counterpart in the comparison data, it is not considered intrinsic. If the frequency, phase and amplitude of the signal, the time base of the observations, and the noise characteristics are exactly the same in both datasets, the decision is obvious. But how shall the general (and typical) case be handled where the peaks and the noise are different in the two time series? What if the readings are taken at different times, as in the case of single-channel photometry? The answer is given by the conditional sig, a novel approach to an old problem. Relying on {\sc SigSpec}, it inherits the substantial advantage of unbiased statistical methodology.

An application of {\sc Cinderella} comparing two datasets is presented in Sect.\,\ref{s3.1} below.

Multi-object photometry monitoring three or more objects in one run builds up a scenario where more information is potentially available than can be handled by the procedure outlined above. If more than one constant star is in the observed sample, the comparison of target data with several other time series at once is desired. As mentioned in Sect.\,\ref{s2.3}, this may be achieved by a pairwise comparison of the target dataset vs.~each comparison dataset. Then the arithmetic mean and rms error over all the results provide good estimators for the overall reliability of a peak in the target spectrum.

\subsection{Single comparison dataset}\label{s3.1}

\begin{figure}\includegraphics[width=256pt]{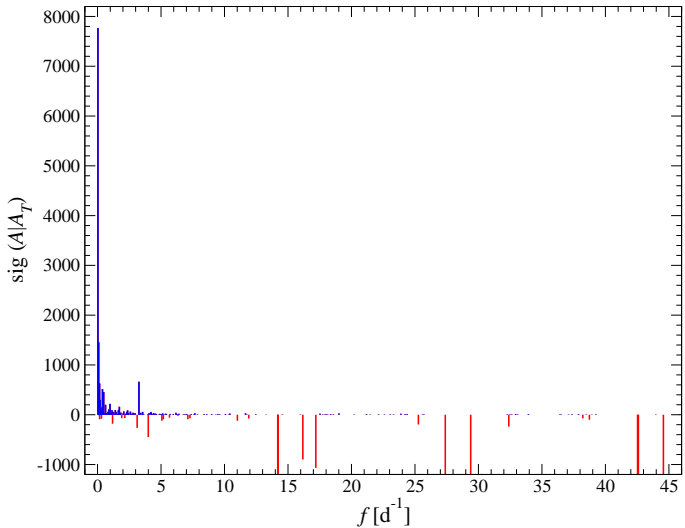}
\caption{{\sc Cinderella} result for $\beta$\,CMi. The conditional sig was estimated for each peak, referring to the background pixel with the highest mean sig between $0$ and $50$\,d$^{-1}$ as the comparison dataset. Blue bars indicate frequencies with $\mathrm{sig}\left( A\left|\right. A_T\right)>5$. The red bars represent frequencies also found in the comparison dataset with $\mathrm{sig}\left( A\left|\right. A_T\right)\le5$.}\label{fig3}
\end{figure}

The first sample scenario concerns MOST measurements of the target star \object{$\beta$\,CMi} and of the sky background. The target data were reduced according to Reegen et al.~(2006). To obtain a most restrictive estimate, the ``worst'' background pixel was used for comparison: a significance spectrum for the intensities of each pixel over time was calculated, and the mean sig in the range from $0$ to $50$\,d$^{-1}$ was used to determine the appropriate pixel. We picked the one with the highest mean sig. The frequency resolution was applied according to Eq.\,(\ref{eqFRes}) with $z=0.75$.

After a comparison of significant signal components ({\sc SigSpec} output) in both reduced target and sky background data using {\sc Cinderella} (Fig.\,\ref{fig3}), the orbital frequency of the spacecraft ($14.2$\,d$^{-1}$), integer multiples and $1$\,d$^{-1}$ aliases are outstanding with their negative conditional sigs, indicating that these frequencies are present in both datasets and hence to be considered instrumental. In the figure, all peaks with a conditional sig $\mathrm{sig}\left( A\left|\right. A_T\right)>5$ are displayed in blue color, the rest in red. The limit of $5$ corresponds to a probability of $10^5$ for the target peak not to be generated by the same process as the corresponding background peak: in one out of $100\,000$ cases, the signal found in the background data plus white noise would produce DFT amplitude in the target data at least as high as the given one.

Of course, a high conditional sig does not definitely rule out a peak to be instrumental. It only tells that no sufficient indication for a common origin of target and background signal at the examined frequency is found. For example, significant orbit-related frequencies may show up for the MOST data also in the {\sc Cinderella} ouput occasionally. This is likely due to the fact that the target area is contaminated by stray light more severely than the sky background available. For a clear statement on the intrinsic (stellar) nature of suspicious peaks that survive the {\sc Cinderella} procedure, follow-up measurements are indispensible. On the other hand, if there is a peak present in the {\sc Cinderella} output, that has to be ruled out for a good reason, the corresponding conditional sig may safely be used as a threshold and applied to the entire spectrum.

Our technique was successfully applied to several MOST targets: \object{AQ\,Leo} (Gruberbauer et al.~2007), \object{$\gamma$\,Equ} (Gruberbauer et al.~2008), and \object{HR\,1217} (Cameron et al.~2008).

\subsection{Multiple comparison datasets}\label{s3.2}

\begin{figure}\includegraphics[width=256pt]{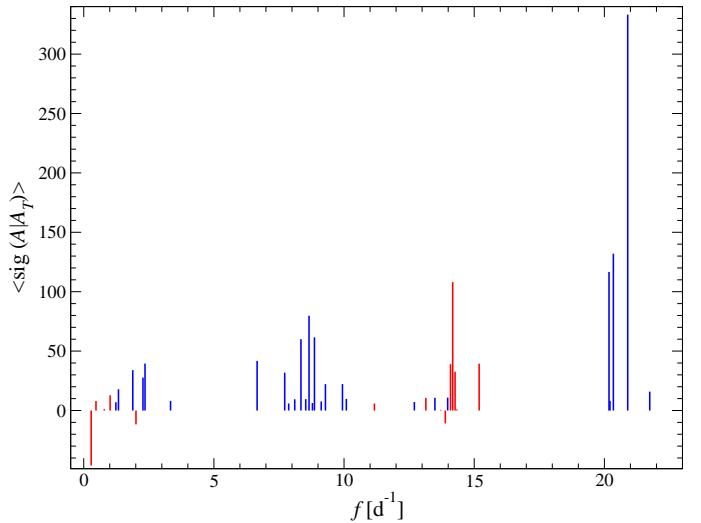}
\caption{{\sc Cinderella} result for HD\,114839. A mean conditional sig was assessed for each peak by averaging over the conditional sigs derived from each individual comparison time series. {\em Blue} bars indicate frequencies with a mean conditional sig exceeding the limit of 5 by more than $3\,\sigma$. Frequencies not meeting this condition are shown in {\em red} color.}\label{fig4}
\end{figure}

In some cases multiple comparison datasets are available. MOST guide star photometry is a good example. While sky measurements are not provided in this observation setup, several light curves of stars which likely suffer from the same contamination by stray light or instrumental trends, are present. However, not every single comparison data set is equally affected and we may not see each instrumental frequency in each DFT spectrum. If we do, the amplitudes (when transformed to some reference mean magnitude value) usually vary from object to object, depending on the position of the stars on the CCD. Still, if these effects are additive in intensities to a first approximation, {\sc Cinderella} provides the means to cope with such a situation due to the statistical nature of the conditional sig. 

The suspected Am star \object{HD\,114839}, a $\gamma$\,Dor/$\delta$\,Sct hybrid observed by MOST using guide star photometry (King et al.~2006), is a good example. It shows intrinsic variability in the low- and intermediate-frequency band, both of which are usually affected by stray light. Since four additional guide stars were observed at the same time, we are able to employ our technique.

In this case, the target dataset was compared to each of the comparison datasets according to the procedure discussed in Sect.\,\ref{s3.1}. The conditional sigs of the four {\sc Cinderella} analyses are averaged, and the standard deviation is computed. These two quantities are used to form a two-fold criterion for the reliability of a target peak. First, a threshold for the conditional sig is defined. In the present example, it is $5$. No peak with a mean conditional sig below this limit is considered intrinsic. Moreover, this threshold has to be exceeded by $k\sigma$, $\sigma$ denoting the standard deviation and $k$ representing an arbitrarily chosen real number. In this case, we use $k=3$. Putting it all together, we only rely on peaks the mean conditional sigs of which exceed $5+3\sigma$.

Fig.\,\ref{fig4} shows the results, which are in very good agreement with King et al.~(2006). It has to be pointed out, however, that in contrast to their method, no manipulation of data other than removal of outliers using $3\,\sigma$ clipping was performed. The blue peaks are considered intrinsic according to the criterion given above. Among the red (rejected) peaks, there are some with $\mathrm{sig}\left( A\left|\right. A_T,\omega ,\theta\right) < 5$ and even negative conditional sigs, but also several peaks where the conditional sigs range up to $100$. In these cases the scatter of sigs in the comparison spectra is very large. Most of the frequencies flagged as artifacts are in the low frequency region below 1\,d$^{-1}$, where nothing survives, and close to the MOST orbit frequency of $14.2$\,d$^{-1}$. In addition, three peaks at $11.2$, $13.2$ and $15.2$\,d$^{-1}$ are rejected, which correspond to $1$\,d$^{-1}$ aliases of the orbit frequency. This aliasing is due to stray light undergoing periodic terrestrial albedo variations as the spacecraft orbits the Earth above the terminator (Reegen et al.~2006).

\section{Composed spectral significances applied to MOST data}\label{s4} 

As described in Sect.\,\ref{s1.4}, the composed sig is a measure of the consistency of a signal detected in multiple data sets, allowing for some mismatch in frequency, amplitude and phase (see also Sect.\,\ref{s3}). This is, for instance, of good use for multisite campaigns, where various instruments with different characteristics are employed. In the case of MOST data, the composed sig can be applied to multiple observing runs on the same star throughout the lifetime of the mission. Significant frequencies consistently detected in multiple data sets will also remain significant in terms of the composed sig. Peaks which are produced by noise will most likely be unique to each observation run. Correspondingly, their composed sig will decrease with increasing number of time series involved.

In the case of conditional sigs, we have one of the involved datasets flagged as target and may search for coincidences using the frequency resolution (Eq.\,(\ref{eqFRes})) about a target frequency. There is no such reference for composed sig computation, because all datasets are considered to be equivalent. Thus we split the frequency range of interest into a sequence of frequency bins. In our example, the grid of bins is ten times finer than the Rayleigh frequency resolution (bin width $\frac{1}{10\,\Delta t}$), and consecutive bins do not overlap. For each bin, the significance spectra for all time series are searched for matching peaks, i.\,e. peaks that either lie in the bin or deviate from it by not more than their Kallinger resolution. If a time series contributes more than one peak to a given bin, only the peak with the highest significance is taken into account. Finally, the composed sig is computed for all peaks associated with the bin.

\begin{figure}\includegraphics[width=256pt]{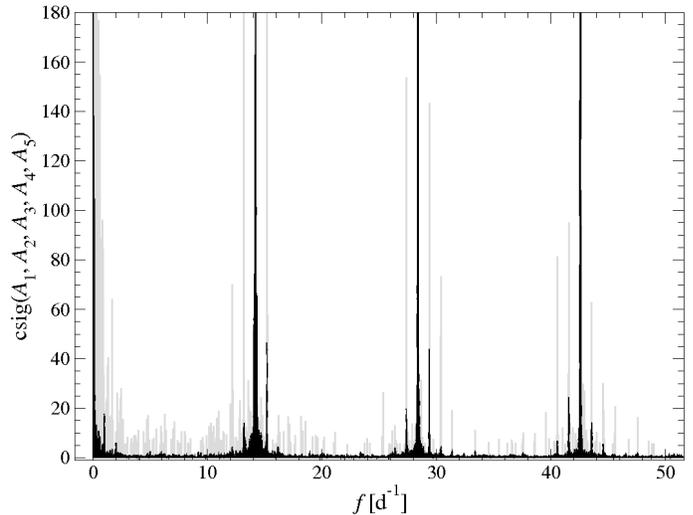}
\caption{A comparison of the composed sig and the individual sig of five background light curves from the 
same observing run. The {\em gray} bars represent an overplot of the significant peaks detected in all five time series individually, as found by {\sc SigSpec}. The {\em black} bars correspond to the results of the composed sig analysis described in Sect.\,\ref{s4}.}
\label{fig:compsig}
\end{figure}

In Fig.\,\ref{fig:compsig} we present the {\sc SigSpec} results of five individual sky background time series from the observing run on the open cluster \object{NGC\,752}. We extracted the sky background signal of five CCD subrasters by selecting pixels which are, to a first approximation, not influenced by any stellar PSF. Each time series was analyzed individually with {\sc SigSpec}. What we expected was that in the individual DFT spectra, the stray-light induced orbit peaks and their $1$\,d$^{-1}$ aliases would be accompanied by spurious peaks at lower sig, whereas the composed sigs would produce a spectrum that would only contain features that referred to long-periodic trends and stray light. The gray graph represents an overplot of all five individual significance spectra. Between the orbit harmonics and their aliases, lots of peaks are visible.

The black plot refers to the composition of all five light curves. Only long-term trends, common to all five datasets, as well as signal corresponding to the orbital frequency of the stray light are considered to be significant. Furthermore, $1$\,d$^{-1}$ sidelobes of the orbital harmonics are visible, referring to daily stray light modulations probably induced by the dependence of the terrestrial albedo on the position over the Earth's surface. Other frequencies, clearly visible in the significance spectra of the individual time-series, are not consistently detected and are therefore regarded as noise. 

\section{Conclusions and outreach}\label{s5} 

This paper introduces a technique to interpret periodicities in an ensemble of data of common origin. {\sc Cinderella} relies on {\sc SigSpec} (Reegen~2007), thus benefitting from a correct employment of the complex phase information in Fourier Space on the one hand and a clean statistical description of interrelation of datasets on the other.

The conditional {\sc Cinderella} mode is based on a quantitative comparison between one target and one or more comparison datasets and returns a measure of the probability (conditional sig) for periodicities identified in the target data 
to be deterministically related (to be `unique') to the target.

The composed {\sc Cinderella} analysis returns a measure of the joint probability (composed sig) that a given periodicity observed in individual datasets -- but with different signal strengths -- is not due to noise. Such datasets could contain, e.g., measurements of the same target in different observing runs or with different instruments (e.g., different filters or simultaneous spectroscopy and photometry).

Our experience (as outlined in our examples in Sect.\,\ref{s3}) confirms that {\sc Cinderella} reliably identifies residual instrumental signal in the MOST data even after a fairly sophisticated data reduction in the time-domain and also provides quantitative arguments to distinguish intrinsic from instrumental signal. 

{\sc Cinderella} is a statistically correct technique replacing what experienced observers achieve based on their ``good feelings'' when evaluating, for example, differential photometry, but, of course, the method is not limited to photometry. It quantitatively determines conditional and composed probabilities for matching peaks in DFT spectra of any kind of datasets containing periodicities. 

\subsection*{Acknowledgements}

PR received financial support from the Fonds zur F\"orderung der wis\-sen\-schaft\-li\-chen Forschung (FWF, projects P14546-PHY, P14984-PHY) Furthermore, it is a pleasure to thank D.\,B.~Guenther (St.~Mary's Univ., Halifax), M.~Hareter, D.~Huber, T.~Kallinger (Univ.~of Vienna), R.~Kusch\-nig, J.\,M. Matthews (UBC, Vancouver), A.\,F.\,J.~Moffat (Univ.~de Montreal), D.~Punz (Univ.~of Vienna), S.\,M. Rucinski (D.~Dunlap Obs., Toronto), D.~Sasselov (Harvard-Smithsonian Center, Cambridge, MA), G.\,A.\,H.~Walker (UBC, Vancouver), and K.~Zwintz (Univ.~of Vienna) for valuable discussion and support with extensive software tests.

Finally, we address our very special thanks to S.\,M.~Mochnacki (University of Toronto) for his careful revision and valuable comments that substantially improved the presentation of this work.

\end{document}